\documentstyle[preprint,aps]{revtex}
\tightenlines
\begin{document}
\draft
\title{Dimensional Crossover in the effective second harmonic 
generation of films of random dielectrics}
\author{P.M. Hui$^{1}$, C. Xu$^{2,1}$, and D. Stroud$^{3}$}
\address{$^{1}$ Department of Physics, The Chinese University of 
Hong Kong, \\
Shatin, New Territories, Hong Kong \\
$^{2}$ Department of Physics, Suzhou University, Suzhou 215006, 
People's Republic of China\\
$^{3}$ Department of Physics, The Ohio State University, Columbus, 
Ohio 43210-1106}
\maketitle
\begin{abstract}
The effective nonlinear response of films of random composites
consisting of a binary composite with nonlinear particles randomly
embedded in a linear host is theoretically and numerically studied.
A theoretical expression for the effective second harmonic
generation susceptibility, incorporating the thickness of
the film, is obtained by combining a modified
effective-medium approximation with the general expression
for the effective second harmonic generation susceptibility
in a composite.  The validity of the theoretical results 
is tested against results obtained by 
numerical simulations on random resistor networks.
Numerical results are found to be well described by our theory. 
The result implies that the effective-medium approximation
provides a convenient way for the estimation of
the nonlinear response in films of random dielectrics.
\end{abstract}

\newpage
\section{Introduction}
Nonlinear response of random composite materials has been a subject
of intensive research for many years\cite{shalaev1}.
The nonlinear response in such composites may be affected by
several factors.  For example, the response can be greatly enhanced
by local field effects, and the percolation effect could change the
character of the system
\cite{stroud1,neeves,hui1,stroud2,aharony,haus,levy1,shalaev2}.
In particular, enhancement of 
nonlinear optical response in random dielectrics may be useful 
for the design of novel switching devices in photonics 
and real-time coherent optical signal processors.  
A number of theoretical works have been developed specifically 
to weakly nonlinear composites
\cite{stroud1,neeves,hui1,stroud2,aharony,haus,levy1,shalaev2,zeng1,hui2,bergman,sipe}.
An excellent review on the recent developments of the subject 
has recently been given by Shalaev\cite{shalaev3}. 
 
While most of the work have focused on nonlinear effects such as
the Kerr effect, Hui {\em et al}\cite{hui3,hui4} 
have derived general expressions
for the effective susceptibilities for second harmonic 
generation (SHG) and third harmonic generation (THG) in a binary 
composite of random dielectrics. In the
dilute limit, the effective SHG and THG
susceptibilities obtained in Refs.\cite{hui3} and \cite{hui4} 
reduce to earlier expressions found by Levy {\em et al.}\cite{levy2}.
For many applications and materials fabrication, samples in the
form of thin films are essential. It is, therefore, useful to study
the effective SHG in films of random dielectrics. 
In the present work, we consider films of random dielectrics
consisting of spherical particles with nonlinear response embedded
randomly in a linear host.  We study the
dependence of the effective SHG on the film thickness
and the concentration of the nonlinear component.
By invoking a modified effective-medium approximation (EMA) to
incorporate the film thickness in the estimation 
of the local fields, together with our previous general expression 
for SHG susceptibility, the 
effective SHG susceptibility including the effect of film thickness
can be calculated. 
The validity of our theory is tested against results obtained by
numerical simulations on random nonlinear resistor networks of different
thickness consisting of two different kinds of resistors.  
It is found that our EMA provides a good description of numerical 
results for the whole range of thicknesses corresponding to systems 
spanning from two dimensions to three dimensions.

The plan of the paper is as follows.  In Sec.II, we present our formalism
on the effective SHG susceptibility of films of random dielectrics.  Section
III gives a description on the model of numerical simulations.  Results
of numerical simulations are compared with theoretical results in Sec.IV. 

\section{Effective medium approximation for SHG in films}
We consider a macroscopically inhomogeneous medium 
consisting of a random mixture of 
two types of materials.  The materials $a$ and $b$ 
have different macroscopic {\bf D}-{\bf E} relations. 
If we only include the quadratic nonlinearities, the 
general form of the {\bf D}-{\bf E} relationship 
at zero frequency would be
\begin{eqnarray}
D_i=\sum_{j} \epsilon_{ij} E_j+\sum_{jk}d_{ijk}E_jE_k, \hspace{1 cm} i=x,y,z, 
\end{eqnarray}
where $D_i$ ($E_i$) is the $i$th component of the 
displacement (electric) field {\bf D} ({\bf E}). For simplicity, 
we consider a composite in which a volume fraction $p$ of nonlinear 
material $a$ is mixed with a volume fraction $1 - p$ of 
linear material $b$, i.e., $d_{ijk}({\bf r}) = d_{ijk}^{a}$ for 
regions occupied by material $a$ and $d_{ijk}({\bf r}) = 0$ for 
regions occupied by material $b$.  We also assume that 
the linear dielectric constants of both components are isotropic, i.e., 
$\epsilon_{ij} = \epsilon^{\alpha} \delta_{ij}$, with 
$\alpha$ = $a$ ($b$) for regions occupied by material $a$ ($b$). 
In general, 
when a monochromatic external field is applied, 
the nonlinearity of the component inside the composite 
will generate local potentials and fields at all harmonic 
frequencies. Here, we focus on the effective 
second harmonic generation (SHG) for a thin of thickness $L$.  

For a finite frequency external applied field of the form
\begin{equation}
E_0(t)=E_{0,\omega}e^{-i\omega t}+c.c.,
\end{equation}
the effective SHG susceptibility can be extracted by considering 
the volume average of the displacement field at the frequency $2\omega$ 
in the inhomogeneous composite medium.  
A general expression for the effective SHG susceptibility
has been derived in Ref.\cite{hui3}.  The result
is that to first order in the SHG susceptibility $d_{ijk}$, 
the effective SHG susceptibility  
is given by\cite{hui3}
\begin{equation}
(d^e_{-2\omega;\omega,\omega})_{ijk}
=<K_{2\omega;il}d_{lmn}K^T_{\omega,mj}K^T_{\omega,nk}>,
\end{equation}
where $<\cdots>$ denotes a volume average over the 
inhomogeneous medium.  
The tensors $K$ and $K^T$ are possible local field enhancement
factors of the forms
\begin{eqnarray}
K_{2\omega;il}({\bf x})=\frac{E_{2\omega;l}({\bf x})}{E_{0,2\omega;i}}, 
\end{eqnarray}
and
\begin{eqnarray}
K^T_{\omega;mj}({\bf x})=\frac{E_{\omega;m}({\bf x})}{E_{0,\omega;j}},
\end{eqnarray}
with a similar expression for $K^T_{\omega;nk}({\bf x})$.  
The tensor $K^T_{\omega;mj}({\bf x})$, for example, is the induced 
$m$th Cartesian component of the electric field at position ${\bf x}$ 
and frequency $\omega$ when a field $E_{0,\omega}$ is applied in the 
$j$-th direction at the same frequency in a {\em linear} random medium 
of the same spatial inhomogeneous structure.   Similar 
expression for the effective THG susceptibility can also be 
derived\cite{hui4}. 

In a composite in which 
only the material $a$ is nonlinear, the effective SHG susceptibility
takes on the simple form 
\begin{equation}
{\bf d}^e_{-2\omega;\omega,\omega}
=p\frac{<E^a_{2\omega}{\bf d}^a_{-2\omega;\omega,\omega}
E^a_\omega E^a_\omega>}{E_{0,2\omega}E_{0,\omega}E_{0,\omega}}.
\end{equation}
It is desirable to further develop a simple analytic approximation
for the average of the local field factors in Eq.(6).
One such approximation is 
the effective medium approximation (EMA).   One possible
effective medium approximation is to calculate
the electric fields within each particle as if that particle
is embedded in an effective medium with dielectric
constant $\epsilon^{e}$\cite{hui4}.  In this approximation, Eq.(6) becomes
\begin{equation}
{\bf d}^e_{-2\omega;\omega,\omega}=p {\bf d}^a_{-2\omega;\omega,\omega}
(\frac{D\epsilon^e_{2\omega}}
{\epsilon^a_{2\omega}+(D-1)\epsilon^e_{2\omega}})
(\frac{D\epsilon^e_{\omega}}
{\epsilon^a_{\omega}+(D-1)\epsilon^e_{\omega}})^2,
\end{equation}
where $D$ is the spatial dimension and the subscripts $\omega$ 
and $2\omega$ specify the frequency at which the local field 
is evaluated.

The effective linear dielectric constants
$\epsilon^{e}_{\omega}$ and $\epsilon^{e}_{2\omega}$ can be
evaluated from the usual EMA for linear random composites by 
\begin{equation}
\sum_{\alpha=a,b}p_\alpha(\frac{\epsilon^\alpha-\epsilon^e}
{\epsilon^\alpha+(D-1)\epsilon^e})=0,  
\end{equation}
where $p_\alpha$ is the volume fraction of the $\alpha$th component and
the summation is over the different components in the composite. 
For a thin film of thickness $L$, the behaviour of the system 
lies somewhere between $D=2$ for small thickness 
and $D=3$ for large thickness.  To incorporate the effects of 
finite thicknesses, it is most convenient to convert the continuum 
description into a discrete version of the problem.  In the discrete 
version, the system is considered to be a lattice consisting of 
$N\times N \times L$ sites, where $L$ is the number of layers.  
For a $D$-dimensional hypercubic lattice, the coordination 
number $z$, i.e., the number of nearest neighbors, 
is related to $D$ through $D = z/2$.  For a single layer ($L=1$), 
$z=4$ and for a large number of layers ($L \rightarrow \infty$), 
the system is three dimensional with $z=6$.  For a system of   
finite thickness $L$, the average coordination number becomes 
$L$-dependent and takes on 
\begin{equation}
z(L)=2(3-\frac{1}{L}). 
\end{equation}
Incorporating Eq.(9) into Eqs.(7) and (8) by replacing 
the dimensionality $D$ by $z(L)/2$ 
constitute a set of equations for a modified 
EMA for the effective SHG susceptibility of films of random 
dielectrics.

\section{Numerical simulations using random nonlinear resistor network}
To test the validity of the modified EMA 
for SHG susceptibility in a film of random dielectrics, we follow 
the approach in Ref.\cite{hui4} and perform 
numerical simulations using a random nonlinear resistor network. 
Such an approach has been applied previously with success 
to study linear optical properties and 
nonlinear properties such as the Kerr effect in random 
composites\cite{zhang,hui5,koss,zeng2}.  
To model a system of finite thickness, two types of bonds are placed 
randomly onto a $N \times N \times L$ lattice, with $L$ running from $1$   
to $N$ modelling a 2D system ($L=1$) and 3D system ($L=N$), respectively.   
In the resistor network model, it is more convenient to define the model 
in terms of conductance of the bonds. 
With probability $p$, the bonds in the lattice 
are occupied by nonlinear components characterized by 
the two parameters: the linear conductance $g^{a}$ 
and the nonlinear susceptibility $d$, with the latter 
representing the response to two voltages at frequency $\omega$.  With 
probability $(1-p)$, the bonds are occupied by linear components 
characterized by a linear conductance $g^{b}$.  To model a mixture of 
metallic and insulating components at finite frequencies, we take 
$g^{a}$ and $g^{b}$ to be of the forms
\begin{equation}
g^a(\omega)=\frac{1+i\omega RC-\omega^2LC}{R+i\omega L},
\end{equation}
and
\begin{equation}
g^b(\omega)=i\omega C.
\end{equation}
The nonlinear susceptibility $d$ of material $a$ is assumed to be 
independent of frequency, for simplicity.  
This choice of $g^a$ and $g^b$ has been widely used in studying the 
frequency response in random composites\cite{hui4,koss,zeng2}, 
because the ratio $g^a/g^b$ has the same form as the Drude 
dielectric function of 
a metal with the plasma frequence $\omega_p=\sqrt{1/LC}$ 
and relaxation time $\tau=L/R$. 
In what follows, the parameters are 
chosen to be $C=1, L=1$, and $R=0.1$, corresponding
to a plasma frequency $\omega_p=1$. 

The effective response of the random nonlinear resistor network 
can be studied by direct numerical simulations\cite{hui4}. 
Using the Kirchhoff's law, the voltage at each node can be solved. 
Although the applied voltage imposed on the nonlinear network is 
taken to consist 
only of a component at frequency $\omega$, the voltage 
at each node inside the network has components at frequencies $\omega$ and $2\omega$. 
So, solving the Kirchhoff's law 
leads to solving two equations $\sum I^{(\omega)}=0$ and 
$\sum I^{(2\omega)}=0$ at each node simultaneously, where the summation 
is over all the bonds connected to the node under consideration and 
$I^{(\omega)}$ and 
$I^{(2\omega)}$ are the component of the current 
corresponding to the frequency $\omega$ and $2\omega$, respectively. 
These two equations at each node 
must be solved self-consistently for the voltages at frequencies $\omega$ and $2\omega$, and 
the effective SHG can be then extracted.  
In what follows, we use a lattice of size 20$\times$20$\times$$L$, with 
an applied voltage set to unity. 
By varying the layer thickness $L$, dimensional crossover of the  
effective SHG response from 2D to 3D can be studied. 

\section{Results and Discussion}

Numerical simulations allow for the extraction of the effective 
linear and nonlinear response of the inhomogeneous system for different 
concentrations 
of the nonlinear component and for different film thicknesses. 
Results from numerical simulations can then be compared with those 
calculated via the modified EMA so as to establish the validity 
of the approximation.  Figure 1 and Figure 2 
show the effective linear response $g_{eff}$ and effective SHG 
response $d_{e}/d$ of the random nonlinear network as a function 
of frequency $\omega$ for three values ($p=0.1, 0.5, 0.9$) of the 
concentration $p$.  In each panel of the figures, the symbols give 
the simulation results and the lines give the EMA results.  
The symbols represent an average over 50 independent runs 
corresponding to different configurations of the 
random network.  The EMA results are obtained by using Eqs.(7)-(9), 
with the dielectric constants replaced by the conductances. 
Four different layer thicknesses are studied for each 
value of the concentration. 

In the dilute case ($p=0.1$), the concentration is lower than 
the percolation threshold regardless of the thickness, i.e., $p<p_{c}$ 
for both $p_{c}(2D)$ and $p_{c}(3D)$ in the bond percolation 
problem.  In this case, isolated nonlinear bonds (component $a$) are 
surrounded by the insulating bonds (component $b$).  In the 
continuum case, the situation corresponds to one in which 
isolated disks (or spheres) of component $a$ are embedded 
in a background of component $b$.  The structure observed 
in $g_{eff}$ corresponds to the so-called surface plasmon resonance 
structure. The surface plasmon resonance peak shifts to lower frequency
as the layer thickness increases.  The shift is related to the 
change in the resonance frequency from $\omega_{p}/\sqrt{3}$ 
for an isolated sphere in 3D (large $L$) to $\omega_{p}/\sqrt{2}$ for 
an isolated disk in 2D ($L=1$).  
The real and imaginary parts of the nonlinear response normalized to the 
SHG coefficient of the nonlinear component $d_{e}/d$ show a corresponding 
resonance structure near the surface plasmon resonance frequency.  
The modified EMA captures the features reasonably well, with a better 
agreement for large values of $L$.  It is reasonable in that 
the EMA, being a mean field approximation, is expected to work 
better in high spatial dimensions in which fluctuation effects 
are smaller.  We also note that the simulation results indicate 
that the system picks up three dimensional character readily 
for not-so-large layer thicknesses.  

For $p=0.5$, which is the percolation threshold for an infinite 2D system 
and higher than $p_{c}(3D)$, the numerical results for $Re(g_{eff})$ with 
$L=1$ show a broad structure with the values at low frequencies tend to 
be enhanced.  For $L\geq 2$, large $Re(g_{eff})$, relative to the dilute 
case, at low frequencies indicates that the composite behaves effectively 
as a conducting (metallic) system.  The greater the thickness, the 
larger is $Re(g_{eff})$.  It is because the value of $p=0.5$ 
becomes increasingly larger than $p_{c}$ as $L$ increases, and the   
system shows increasingly dominating metallic character.  The 
effective SHG response shows a broader structure with an enhanced 
$d_{e}/d$ compared with the dilute concentration case.  The enhancement 
in the SHG response comes about from the enhancement in the local field 
applying on the nonlinear component when the system has more better 
conducting nonlinear component.  The modified EMA, again, gives results 
in reasonable agreement with simulation results, indicating that
incorporating a thickness-dependent coordination number into  
EMA can be used 
to estimate the effective SHG response in random dielectric systems. 

For high concentrations ($p=0.9$) of the nonlinear component, 
the systems shows metallic behavior in the linear response, 
as shown in Fig.1.  The behavior 
of the effective SHG susceptibility basically follows that of the case of 
$p=0.5$, with the appearance of a small resonance structure 
at low frequency. The structure is due to the isolated 
insulating component in an otherwise metallic background. 


In both our numerical and EMA results,
the SHG composite susceptibility is enhanced only relatively
weakly over that of the pure SHG material.  By
contrast, such processes as Kerr nonlinearity
are greatly enhanced in composite 
media\cite{stroud1,neeves,hui1,stroud2,haus,levy1,shalaev2,zeng1,hui2,bergman,sipe,ma,sarychev}.  
We believe this weak enhancement may occure because Eq.\ (6) involves
the linear fields at {\em two different frequencies}.  Thus, the enhancement
will be smaller than when all the field enhancement factors
are resonant at the {\em same} frequency.  The enhancement is further
weakened by randomness when both components are present in high
concentrations, which causes the field enhancement factors to
be broadened over a range of frequencies.  At very low concentrations,
despite the two frequencies, other studies have shown that the SHG enhancement is still large.

Also, our calculations assume that only the metallic component 
has a nonzero SHG susceptibility. We have done a few test EMA calculations
in 2D, and find a somewhat larger enhancement, when only the
nonmetallic component is nonlinear, with Im($d_{eff}/d$)
and Re($d_{eff}/d$) reaching magnitudes as large as 4 or 5.

     
Expression (6) for the
effective SHG susceptibility treats the composite
as a {\em homogeneous medium with effective properties}, as  
is reasonable if the inhomogeneities
are small compared to a wavelength.  Since the medium is effectively
homogeneous, the SHG emission will be in the form of a
collimated beam.  Besides this beam, there
is another SHG contribution, resulting from scattering 
from the inhomogeneities.  This contribution, which has a broad
angular distribution, depends on particle size, unlike that 
calculated here.  It has been seen 
in experiments on inhomogeneous metal-insulator films\cite{breit},
and may be large near the percolation threshold.

\section{Summary}

We have studied the dependence of the effective 
SHG response in a random dielectric system as a function 
of the concentration of the nonlinear constituent and 
the systemn thickness.  A modified EMA,
based on a previous expression 
of the effective SHG derived by the authors coupled with a 
thickness-dependent coordination number, is proposed for the 
effective SHG response.  Numerical simulations on a random 
network consisting of a mixture of nonlinear and linear 
conductances allow extraction of the effective linear 
and SHG response.  It is found that the modified EMA gives 
results that are in reasonable agreement with the numerical
simulation results.  The system picks up 
three dimensional character for moderate film thicknesses. 
The results show that the modified EMA can be used as an 
effective tool for estimating the effective SHG response in 
random dielectric 
systems.  The approximation can be applied to other systems 
consisting of nonlinear components.  For example, the nonlinear 
SHG response may be associated 
with the poorer conductor of the two constituents.  In this case, 
the contrast in the linear dielectric 
constants between the two constituents may produce an enhanced 
local field in the poorer conducting component and thus 
drive a greater SHG enhancement.  Indeed, some preliminary calculations, 
using the EMA, do give rise to this increased enhancement.

\acknowledgments{This work was supported in part by a grant from the 
Research Grants Council of the Hong Kong SAR Government through 
the grant CUHK4129/98P.  One of us (DS) 
acknowledges support from 
the National Science Foundation through the grant DMR01-04987.}

\begin{figure}
\bigskip
\caption{The effective linear response $g_{eff}$ as a function of $\omega$. 
The three panels correspond to $p$ = $0.1$, $0.5$, and $0.9$. 
In each panel, results for thickness $L=1$, $2$, $4$, and $6$ are shown. 
The symbols are results of numerical simulations 
and the curves are theoretical results based on the EMA.}
\bigskip
\bigskip

\caption{The effective SHG response as a function of $\omega$. 
The three panels correspond to $p$ = $0.1$, $0.5$, and $0.9$. 
In each panel, results for thickness $L=1$, $2$, $4$, and $6$ are shown. 
The symbols are results of numerical simulations 
and the curves are theoretical results based on the EMA.}

\end{figure}


\newpage
\begin{thebibliography}{99}
\bibitem{shalaev1} V. M. Shalaev, Phys. Rep. {\bf 272}, 61 (1996);    
D. J. Bergman and D. Stroud, Solid State Physics {\bf 46}, 
147 (1992); 
C. Flytzanis, F. Hache, M. C. Klein, and P. Roussignol, 
in {\it Progress in Optics} {\bf 29}, edited 
by E. Wolf (North-Holland, Amsterdam, 1991), p. 322.
\bibitem{stroud1} D. Stroud and Van E. Wood, 
J. Opt. Soc. Am. B {\bf 6}, 778 (1989). 

\bibitem{neeves} A. E. Neeves and M. H. Birnboin, 
J. Opt. Soc. Am. B {\bf 6}, 787 (1989); 
Y. Q. Li, C. C. Sung, R. Inguva, and C. M. Bowden, {\it ibid}, 
{\bf 6}, 814 (1989); 
J. W. Haus, N. Kalyaniwalla, R. Inguva, M. Bloemer, and 
C. M. Bowden, {\it ibid}, {\bf 6}, 797 (1989).
\bibitem{hui1} P. M. Hui, Phys. Rev. B {\bf 41}, 
1673 (1990); {\bf 49} 15344 (1994).
\bibitem{stroud2} D. Stroud and P. M. Hui, 
Phys. Rev. B {\bf 37}, 8719 (1988).
\bibitem{aharony} A. Aharony, Phys. Rev. Lett. {\bf 58}, 
2726 (1987). 
\bibitem{haus} J. W. Haus, R. Inguva, and 
C. M. Bowden, Phys. Rev. A {\bf 40}, 5729 (1989).
\bibitem{levy1} O. Levy and D. J. Bergman, Phys. Rev. B {\bf 46}, 
7189 (1992).
\bibitem{shalaev2} V. M. Shalaev, E. Y. Poliakov, and 
V. A. Markel, Phys. Rev. B {\bf 53}, 2437 (1996) and 
references therein.
\bibitem{zeng1} X. C. Zeng, D. J. Bergman, P. M. Hui, and 
D. Stroud, Phys. Rev. B {\bf 38}, 10970 (1988).
\bibitem{hui2} P. M. Hui, J. Appl. Phys. {\bf 68}, 3009 (1990).
\bibitem{bergman} R. Levy-Nathansohn and D. J. Bergman, 
J. Appl. Phys. {\bf 77}, 4263 (1995).
\bibitem{sipe} J. E. Sipe and R. W. Boyd, Phys. Rev. A {\bf 46}, 
1614 (1992); R. W. Boyd and J. E. Sipe, 
J. Opt. Soc. Am. B {\bf 11}, 297 (1994).
\bibitem{shalaev3} V. M. Shalaev, 
{\em Nonlinear Optics of Random Media: 
fractal composites and metal-dielectric films} 
(Springer, New York 2000); 
see also V. M. Shalaev (ed.), 
{\em Properties of nanostructured random 
media} (Springer, New York, 2002). 
\bibitem{hui3} P. M. Hui and D. Stroud, 
J. Appl. Phys. {\bf 82}, 4740 (1997).
\bibitem{hui4} P. M. Hui, P. Cheung, and D. Stroud, 
J. Appl. Phys. {\bf 84}, 3451 (1998).
\bibitem{levy2} O. Levy, D. J. Bergman, and D. G. Stroud, 
Phys. Rev. E {\bf 52}, 3184 (1995).
\bibitem{zhang} X. Zhang and D. Stroud, Phys. Rev. 
B {\bf 52}, 2131 (1995).
\bibitem{hui5} P. M. Hui, W. M. V. Wan, and 
K. H. Chung, Phys. Rev. B {\bf 52}, 15867 (1995).
\bibitem{koss} R. S. Koss and D. Stroud, Phys. Rev. B 
{\bf 35}, 9004 (1987).

\bibitem{zeng2} X. C. Zeng, P. M. Hui, and D. Stroud, 
Phys. Rev. B {\bf 39}, 1063 (1989).

\bibitem{ma} H. Ma, R. Xiao and P. Sheng, 
J. Opt. Soc. Amer. {\bf B15}, 1022 (1998).

\bibitem{sarychev} A. K. Sarychev and V. M. Shalaev, 
Physics Reports {\bf 335}, 275 (2000).


\bibitem{breit} M. Breit, V. A. Podolskiy, S. 
Gr\'{e}sillon, G. von Plessen,
J. Feldmann, J. C. Rivoal, P. Gadenne, A. K. Sarychev, and 
V. M. Shalaev, Phys. Rev. {\bf B64}, 125106 (2001).


\end{thebibliography}
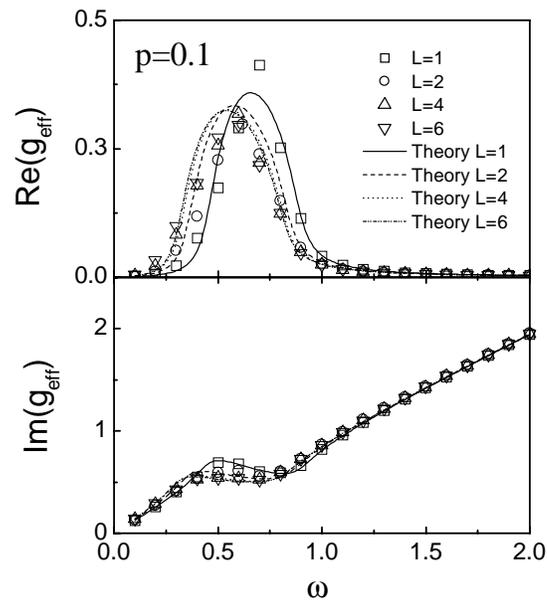
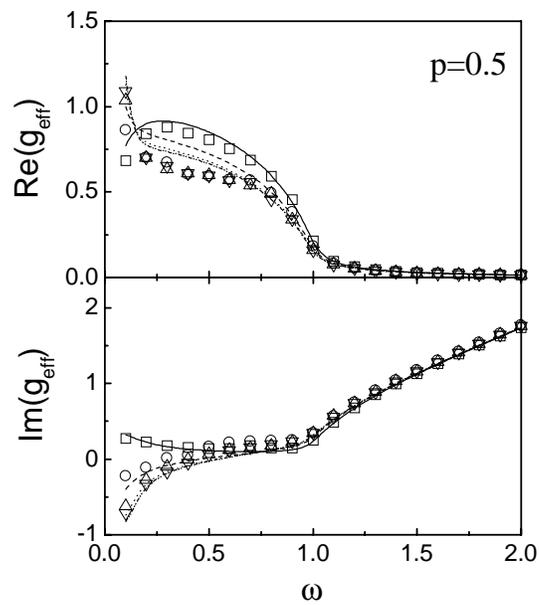
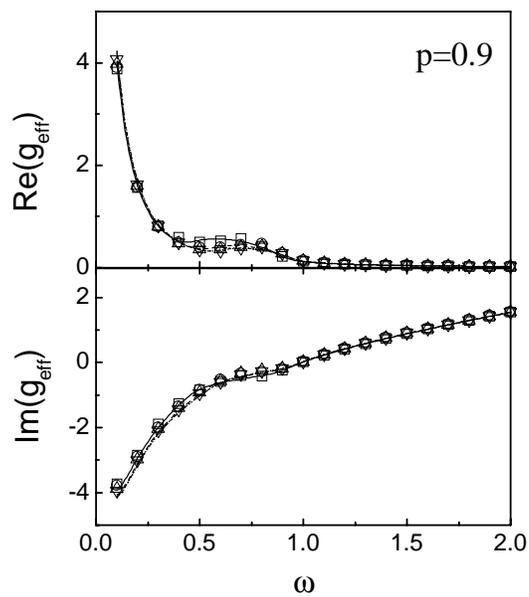
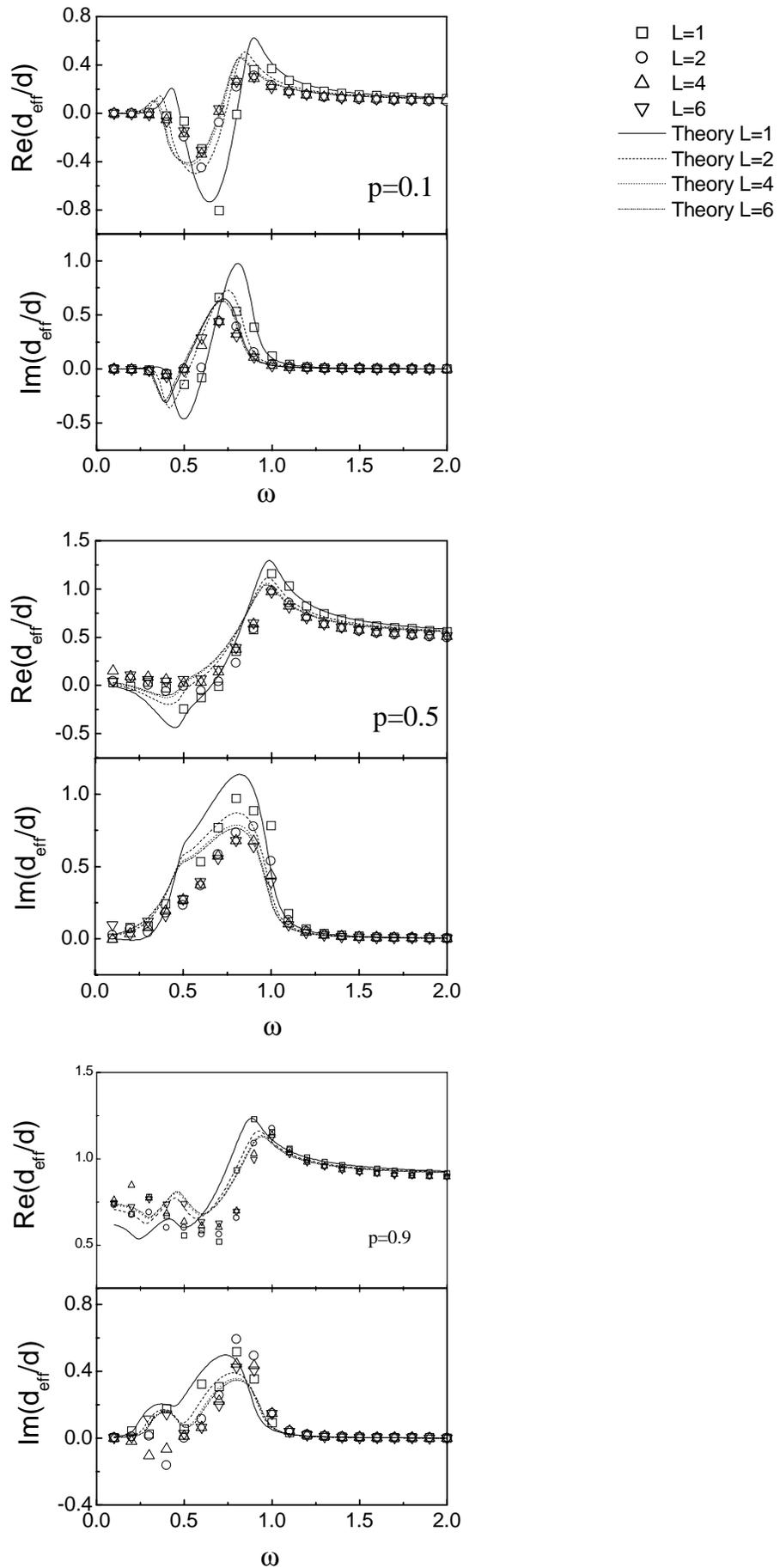
\end{document}